\documentclass[conference]{IEEEtran}
\usepackage{times} 
\usepackage{helvet}
\usepackage{amsmath}
\usepackage{array}
\usepackage{courier} 
\usepackage[noblocks]{authblk} 
\usepackage{graphicx}
\usepackage{color}
\title{A Deep Learning Approach to the Citywide Traffic Accident Risk Prediction}

\author[*]{Honglei Ren}
\author[*]{You Song}
\author[*]{Jingwen Wang}
\author[+]{ Yucheng Hu}
\author[+]{Jinzhi Lei}
\affil[*]{School of Software, Beihang University, Beijing, China,\authorcr songyou@buaa.edu.cn, renhongleiz@buaa.edu.cn,wangjingwen@buaa.edu.cn}
\affil[+]{Zhou Pei-Yuan Center for Applied Mathematics, Tsinghua University, Beijing, China,\authorcr  huyc@tsinghua.edu.cn, jzlei@mail.tsinghua.edu.cn}

\begin{document}
\maketitle

\begin{abstract}
With the rapid development of urbanization, the boom of vehicle numbers has resulted in serious traffic accidents, which led to casualties and huge economic losses. The ability to predict the risk of traffic accident is important in the prevention of the occurrence of accidents and to reduce the damages caused by accidents in a proactive way. However, traffic accident risk prediction with high spatiotemporal resolution is difficult, mainly due to the complex traffic environment, human behavior, and lack of real-time traffic-related data. In this study, we collected big traffic accident data. By analyzing the spatial and temporal patterns of traffic accident frequency, we presented the spatiotemporal correlation of traffic accidents. Based on the patterns we found in analysis, we proposed a high accurate deep learning model  based on recurrent neural network toward the prediction of traffic accident risk. The predictive accident risk can be potential applied to the traffic accident warning system. The proposed method can be integrated into an intelligent traffic control system toward a more reasonable traffic prediction and command organization.
\end{abstract}

\section{Introduction}
In modern society, the rapid development of urbanization has resulted in the boom of vehicles, causing a number of problems, such as traffic congestion, air pollution, and traffic accidents. These problems have caused huge economic loss as well as human casualties. According to \emph{Global Status Report on Road Safety}, published by World Health Organization in 2015, about 1.25 million people were killed in traffic accidents every year. With the help of big traffic data and deep learning, real-time traffic flow prediction has enabled people to avoid traffic jam by choosing less congested routes. Big traffic data and deep learning may also provide a promising solution to predict or reduce the risk of traffic accidents.

One important task in traffic accident prevention is to build an effective traffic accident risk prediction system. If the traffic accident risk in a certain region can be predicted, we can disseminate this information to the nearby drivers to alert them or make them choose a less hazardous road. However, accurate prediction of traffic accident risk is very difficult because many related factors could affect traffic accident. For example, different regions have tremendous difference on traffic accident rate. In addition, poor weather condition such as snow or fog can reduce road visibility and traffic capacity, thus increase the change of traffic accidents. Traffic accident rate varies at different time of a day, possibly related to the physical condition of the drivers. Although many researchers have focused on the identification of key factors associated with traffic accident \cite{Zhang:2013ej}, effective prediction of the traffic accident risk dynamically remains to be a challenge problem.
  
With the development of deep learning, methods based on deep learning and big data have shown favorable results in traffic related problems, such as traffic flow prediction \cite{wu2016short}, arrival time estimation \cite{bai2015dynamic}, origin-destination forecasting \cite{toque2016forecasting}, etc. As for traffic accident risk prediction based on deep learning, to our best knowledge, the only work is done by Chen et. al., who use human mobility features extracted from Stack denoise Autoencoder to infer traffic accident risk in Japan \cite{chen2016learning}. However, they did not consider the periodical patterns and the spatial distribution patterns of traffic accidents. In particular, traffic accidents may closely related to the day of week. Other important factors they missed are weather condition, air quality, etc. To improve the power of traffic accident risk prediction, it is important to combine all these factors into a comprehensive  model.

\begin{figure*}[t]
\centering
\includegraphics[width=6.5 in]{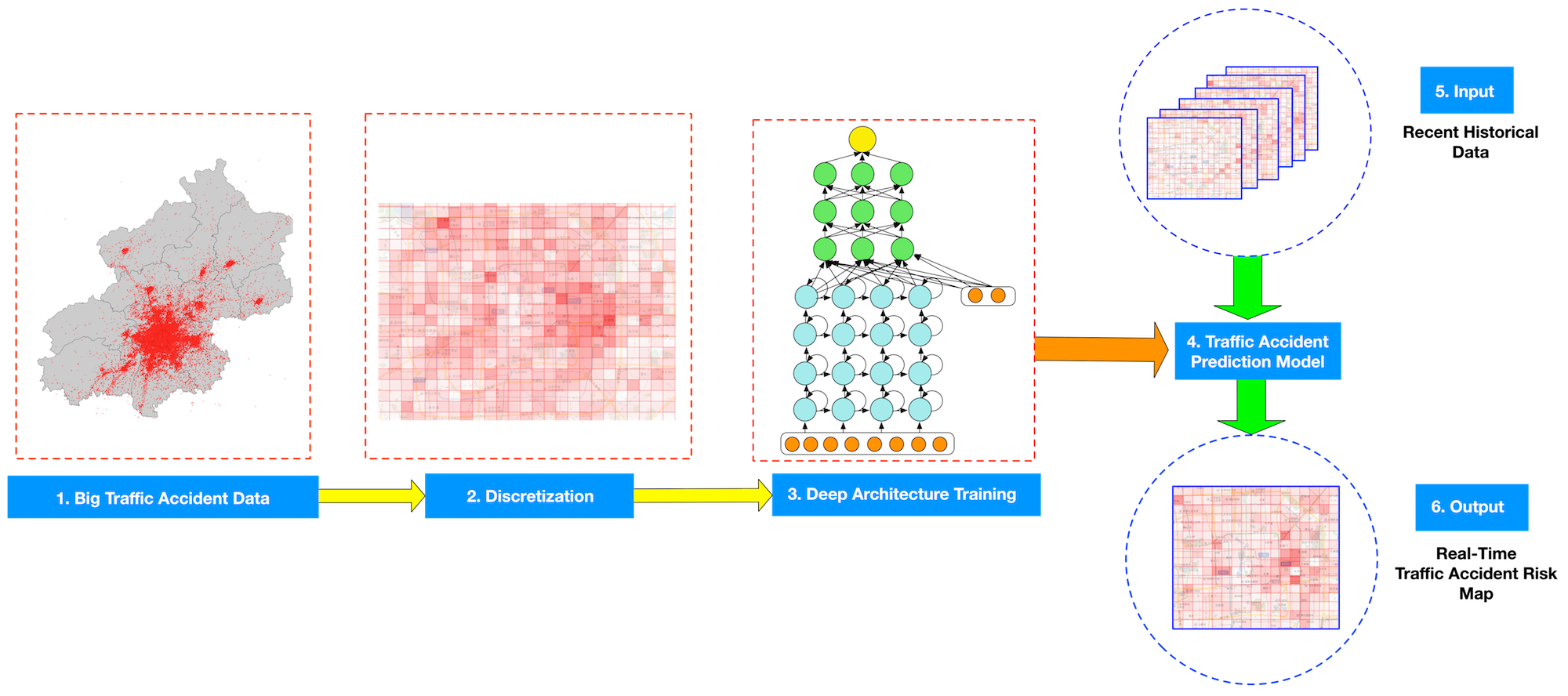}
\caption{Workflow of our traffic accident risk prediction method. First, the big traffic accident data is collected. Second, the data is discretized in space and time, and then feed to the deep model for training. After training, we feed recent historical data to the trained model and then obtained the real-time traffic accident risk prediction.}
\label{fig1}
\end{figure*}

In this paper, we collected big traffic accident data and built a deep model for traffic accident risk prediction based on recurrent neural network. By analyzing the spatial and temporal patterns of traffic accident frequency, we presented the spatiotemporal correlation of traffic accidents. Based on the patterns we found in analysis, we proposed a high accurate deep learning model for traffic accident risk prediction. The model can learn deep connections between traffic accidents and its spatial-temporal patterns. As a potential application, the traffic accident prediction system based on our method can be used to help traffic enforcement department to allocate police forces in advance of traffic accidents.

The rest of this paper is organized as follows: Section 2 introduces some previous works that are related with the present one. Section 3 describes the data source and the pattern analysis result of traffic accidents. Section 4 introduces our deep learning model for traffic accident risk prediction. Section 5 shows the results of experiment. Section 6 gives the conclusions and future works.

\section{Related Work}

\subsubsection{Identification of Traffic Accident Trigger}
Tremendous efforts have been devoted to the identification of key conditions or particular traffic patterns that could lead to traffic accident. For instance, Oh proposed the assumption that disruptive traffic flow is a trigger to crash \cite{oh2001real}. Based on the loop detector data and crash data, they found that 5-min standard deviation of speeds right before a traffic accident is an effective indicator of crash. Although different crash indicators have been proposed, they could not meet the requirement of accurate accident prediction because numerous factors have complex connections with traffic accidents.

\subsubsection{Real-time Traffic Accident Prediction}
With the development of machine learning, many researchers start to focus on real-time traffic accident prediction. Lv chose feature variables based on Euclidean metric and utilized k-nearest neighbor method to predict traffic accident \cite{lv2009real}.Park collected big traffic accident data of highway in Seoul and build a prediction workflow based on k-means cluster analysis and logistic regression  \cite{park2016highway}. Recently, Chen used human mobility data in Japan and build a Stack denoise Autoencoder to infer the real-time traffic risk \cite{chen2016learning}. One limitation of these works is that, they did not incorporate several importance factors such as traffic flow, weather condition, air quality into their model. Without these information, the predictive power of the model could be weakened.

\subsubsection{Deep Learning}
The success of deep learning has proved its power in discovering intricate structures in high-dimensional data. It has been widely used as the state-of-the-art technique in image recognition speech recognition, natural language understanding , etc. As for researches on intelligent transportation system, a number of studies focus on traffic flow prediction based on deep learning \cite{wu2016short}. In a longer time scale, some studies try to predict the congestion evolution of large-scale transportation network \cite{ma2015large}. Another interesting application utilized deep reinforcement learning to control the timing of traffic signal \cite{li2016traffic}.

\section{Pattern Analysis of Traffic Accident}

\subsection{Big Traffic Accident Data}

In this study, to predict traffic accident risk, the traffic accident records of Beijing in 2016 and 2017 was collected. Each record contains the time, GPS (Global Positioning System) coordinate of the accident event.

\subsection{Data Preprocessing}
Before we analyze the pattern of accident, and build machine learning model, a proper data structure is necessary. Therefore, we first preprocess our raw data by discretization. 

The traffic accident data was first discretized in space and time. The temporal resolution was 1 hour for different time horizon of prediction, and spatial resolution dimension was 1000m$\times$1000m in uniform grids. 

After discretization, we obtained a matrix $S$ whose element $S_{r,t}$ is the count of traffic accidents happened within region $r$ and time slot $t$. 

\subsection{Spatial Distribution of Traffic Accident}
\begin{figure}[htbp]
\centering
\includegraphics[width=2.0 in]{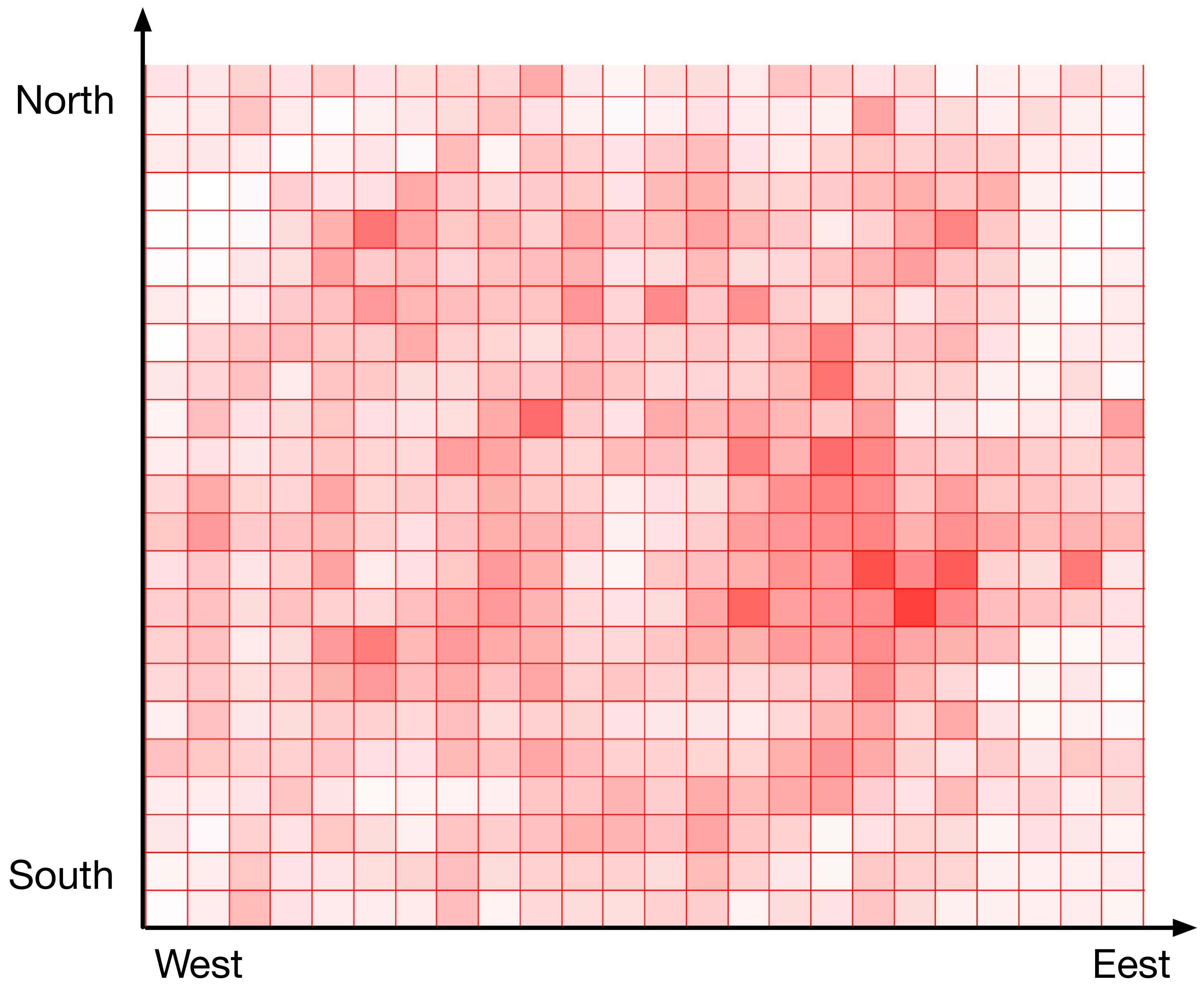}
\caption{The heatmap of traffic accident frequency in Beijing in 2016 with 1000m*1000m spatial resolution. Deeper red indicates higher frequencies of traffic accident.}
\label{fig:spatial}
\end{figure}

To explore whether traffic accident frequency is associated with the geographical position of a region, we plot the heatmap of traffic accident frequency in Beijing in 2016 (Figure \ref{fig:spatial}). As shown in Figure \ref{fig:spatial}, the traffic accident frequency is not uniform distributed, and it is highly related with the geographical position of a region. Usually, the highest traffic accident region lies in the major commercial and business areas.

\subsection{Temporal Pattern of Traffic Accident}
\begin{figure}[htbp]
\centering
\includegraphics[width=3.5 in]{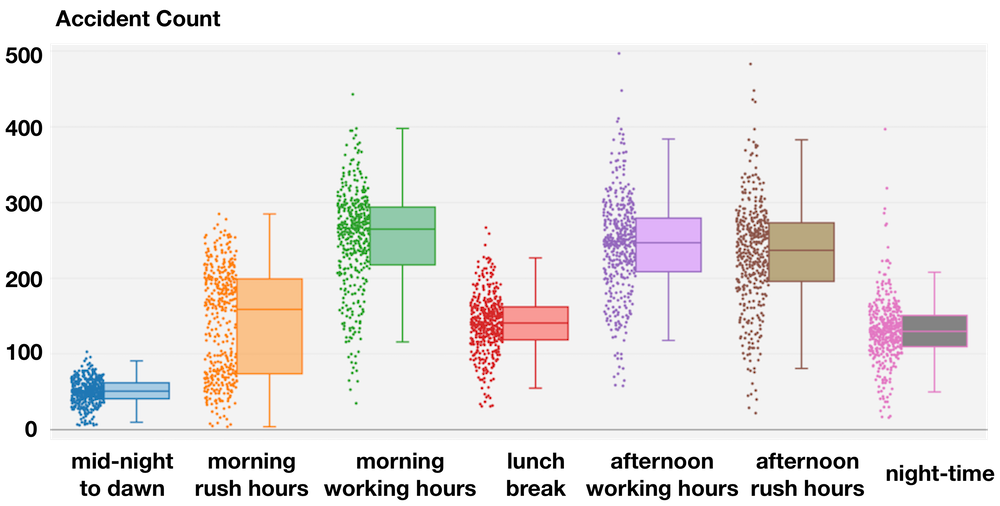}
\caption{Scatter and box-plot of everyday's traffic accident count for different time periods in Beijing.}
\label{fig_temp_daily}
\end{figure}

To explore the temporal patterns of the traffic accident frequency, we first checked whether everyday's traffic accident count varies in different time period. Figure \ref{fig_temp_daily} gives the scatter and box-plot of the everyday's traffic accident count for different time periods of Beijing. Obviously, the traffic accident patterns change drastically for different time period of a day. Specifically, traffic accident is more frequent at rush hours than that at off-peaks.

The time periods in Figure \ref{fig_temp_daily} is defined according to working time pattern and Chinese lifestyle \cite{Zhang:2013ej}: 00:00--06:59 (mid-night to dawn), 07:00--08:59 (morning rush hours), 09:00--11:59 (morning working hours), 12:00--13:59(lunch break), 14:00--16:59 (afternoon working hours), 17:00--19:59 (afternoon rushing hours), and 20:00--23:59 (nighttime).

\begin{figure*}[t]
\centering
\includegraphics[width=6.5 in]{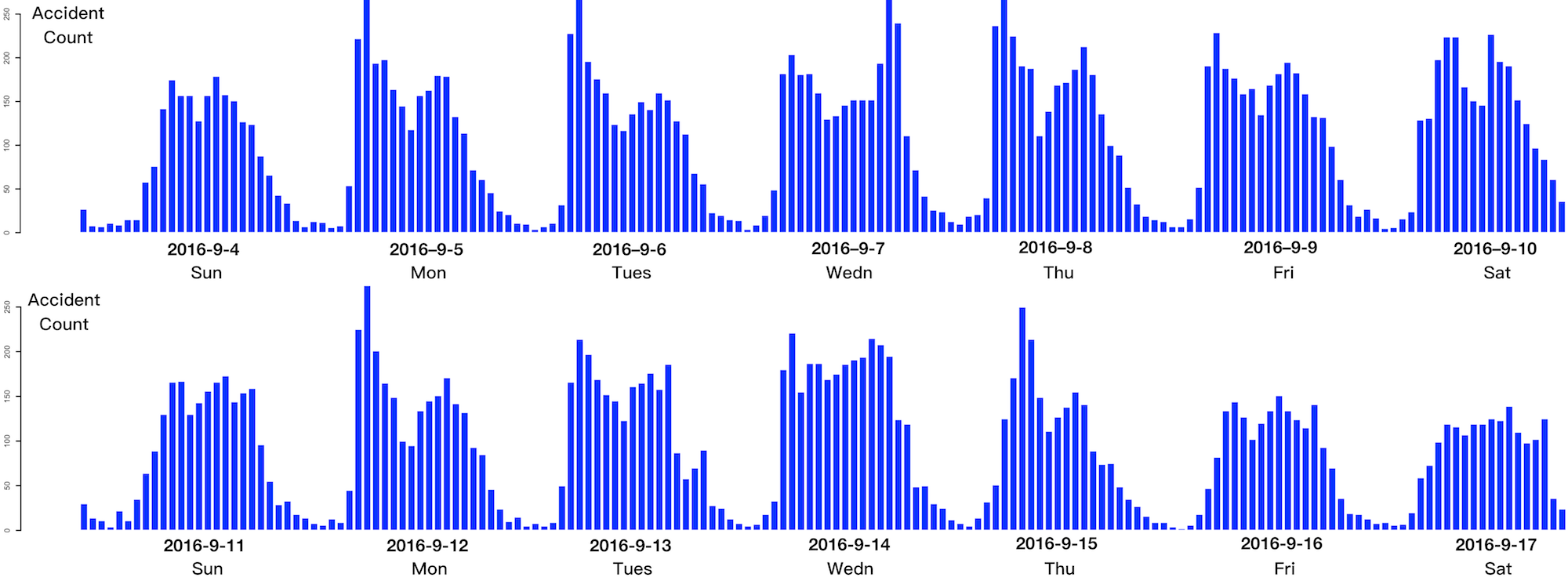}
\caption{Histogram of hourly traffic accident count from 2016-09-04 to 2016-09-17 (two weeks)}
\label{fig_temp_week}
\end{figure*}

Beside temporal patterns for different time period, we also want to know whether weekly periodic patterns exist in traffic accident frequency. Therefore, we plot a two week's histogram of hourly traffic accident count (Figure \ref{fig_temp_week}). It can be observed that the patterns of histograms are similar for the same day of week and between weekdays. 

To quantify the spatio-temporal correlation between traffic accident, we first defined the spatial correlation for a given time $t$ as follows: 

\begin{equation}
C(k,t) = \frac{\sum_{i,j}(a_{i,j,t} - \overline{a}_t)(a_{i',j',t} -  \overline{a}_t)}{\sum_{i,j}(a_{i,j,t} - \overline{a}_t)^2}
\label{eq:spatial_corr}
\end{equation}

where $$\overline{a}_t = \frac{\sum_{i,j}{a_{i,j,t}}}{M * N}$$

The $C(k,t)$ in Eq.(\ref{eq:spatial_corr}) is the spatial correlation with a $k$ Manhattan distance for a given time $t$. $a_{i,j,t}$ is the traffic accident count happened in grid $(i, j)$ and time $t$. $\overline{a}_t$ is the average traffic accident count of all grids at time $t$. $M$ and $N$ are the number of grids along the longitude and latitude.

Based on the spatial correlation defined by Eq.(\ref{eq:spatial_corr}), the spatio-temporal correlation can be written as Eq.(\ref{eq:spatio_temporal_corr}).

\begin{equation}
f(k,\tau) = \frac{\sum_t(C(k,t) - \overline{C}(k))(C(k,t+\tau) - \overline{C}(k))}{\sum_t(C(k,t) - \overline{C}(k))^2}
\label{eq:spatio_temporal_corr}
\end{equation}

where $f(k,\tau)$ is the correlation of two grids with a $k$ Manhattan distance and time interval $\tau$.

\begin{figure}[htbp]
\centering
\includegraphics[width=3.0 in]{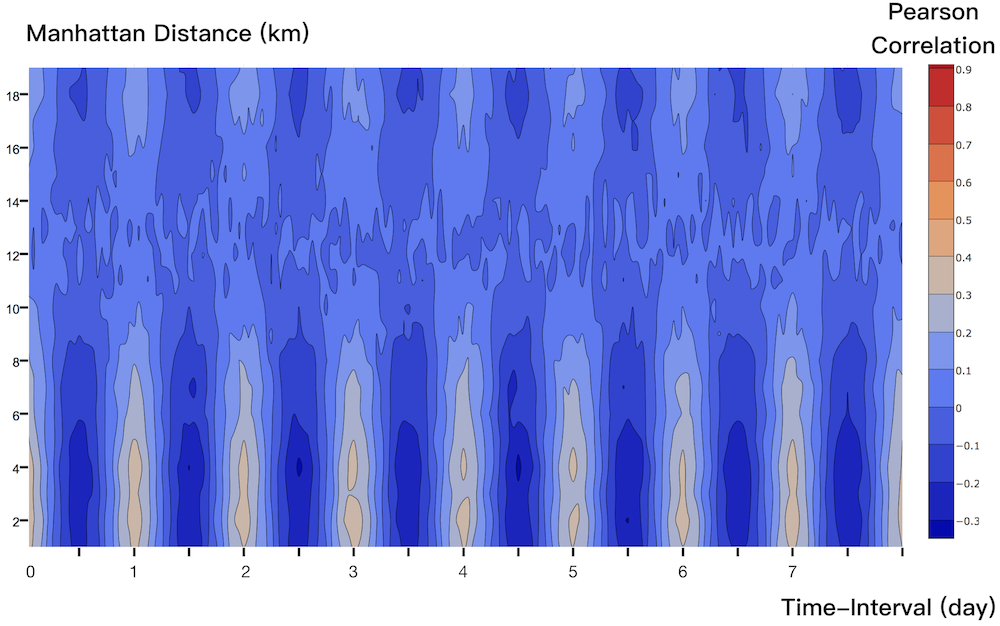}
\caption{The contour map of the spatio-temporal correlation of traffic accident. The horizontal and vertical axis are the time-interval and Manhattan distance of two grids, respectively.}
\label{fig_spatiao_temp_corr}
\end{figure}

Figure \ref{fig_spatiao_temp_corr} shows the contour map of the spatio-temporal correlation of traffic accident. It can be observed that the correlation shows a strong temporal periodic pattern, and the period is around 24 hours. Traffic accidents have about 0.4 $\sim$ 0.5 correlation if their Manhattan distance is within 4 km, and time interval is the multiples of 24 hours.

\section{Deep Model of Traffic Accident Risk Prediction}
\label{sec:deep_model}

\begin{figure}[htbp]
\centering
\includegraphics[width=2.0 in]{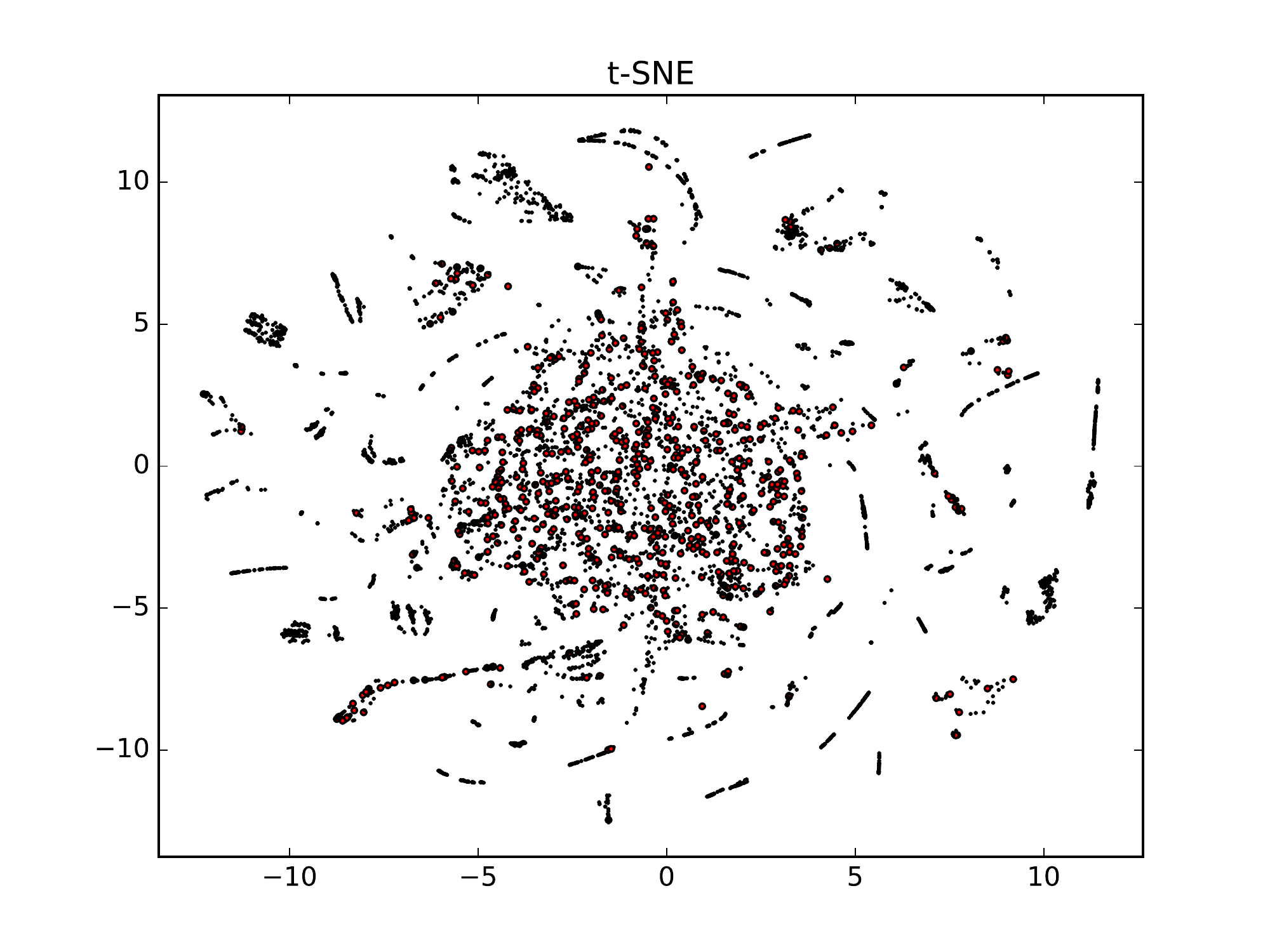}
\caption{Dimensionality reduction and visualization with t-SNE, the original data is the sequence of traffic accident count for a region. The red points are the data with traffic accident, and the black points are the traffic accident free data}
\label{fig_tsne}
\end{figure}

As Chen et. al. has documented, after some analysis of traffic accident data, we find it is difficult to predict whether traffic accident will happen or not directly, because complex factors can affect traffic accident, and some factors, such as the distraction of drivers, can not be observed and collected in advance \cite{chen2016learning}. Figure \ref{fig_tsne} gives the dimensionality reduction result by t-SNE when we predict whether traffic accident happen or not directly, the input data is the sequence of traffic accident count for a region. 
Obviously, the red points (accident) and black points (non-accident) are inseparable, and that means it is hard to predict whether traffic accident happen or not directly. Therefore, we try to predict traffic accident frequency(risk), that is average traffic accident count per hour for the same time of recent days (3 days, 7 days, 30 days, etc.). For instance, if 5 accidents happened during 8:00-9:00 a.m. in last 3 days, then today's traffic accident frequency during 8:00-9:00 a.m is $ \frac{5\ times}{3\ hours} \approx 1.67\ times/ hour$.

Our method is illustrated in Figure \ref{fig1}. First, we discretized the big traffic accident data in space and time, so that it can be processed by machine learning algorithm. Then we constructed a deep model on the basis of recurrent neural network to infer traffic accident risk, and input the processed data into it. After the data training, we input the recent traffic accident frequency data into the trained model, and then obtained the predicted accident risk map from the output.

\begin{figure}[!t]
\centering
\includegraphics[width=2.5 in]{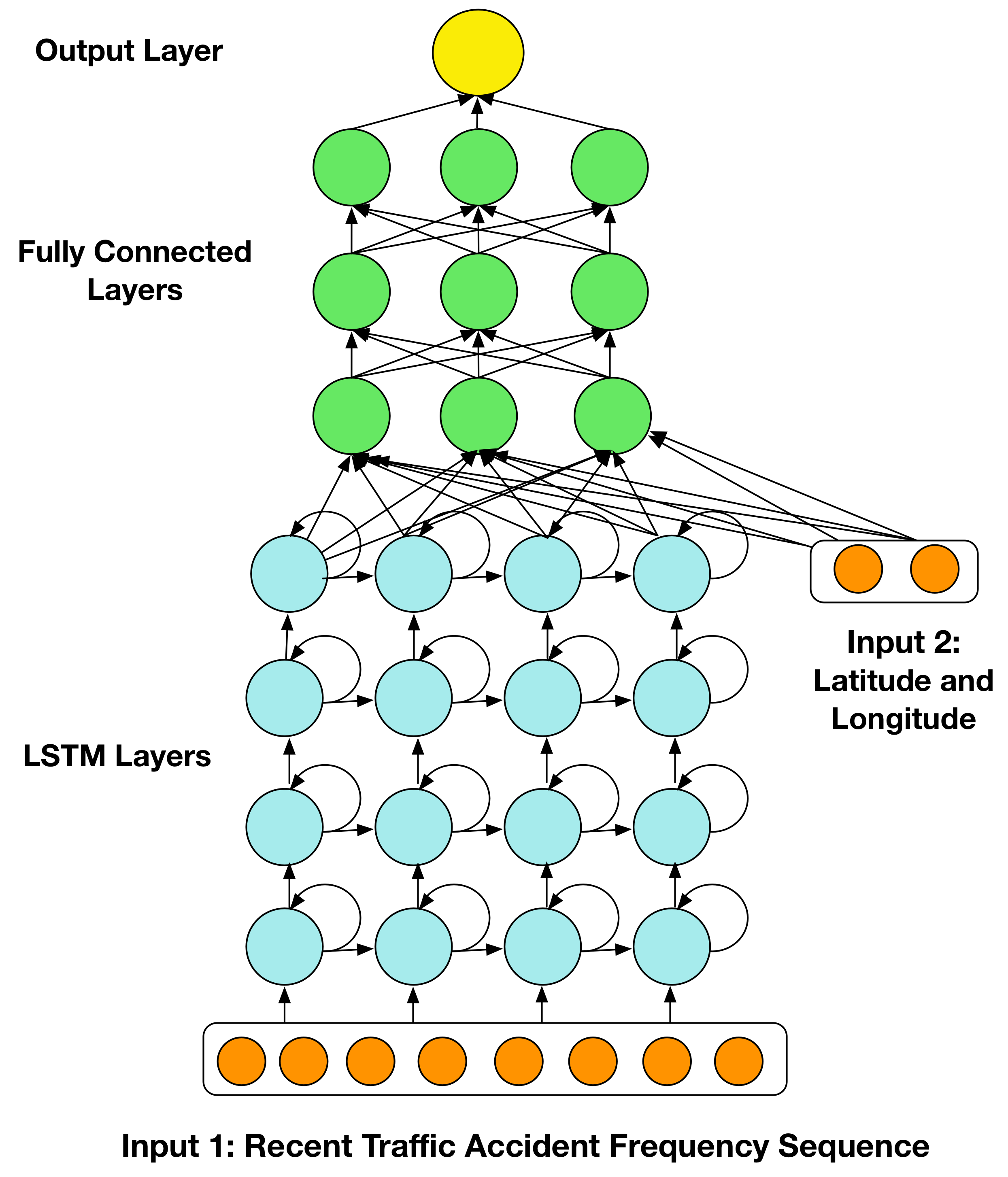}
\caption{The Deep Model of Traffic Accident Risk Prediction Method based on LSTM. The model consisted of 2 separated input layers, 4 LSTM layers, 3 fully connected layers and 1 output layer. The first input layer is made up of the sequence of recent traffic accident frequency. The second input layer contains the longitude and latitude of the region center that we expected to predict.}
\label{fig:model}
\end{figure}

\subsection{Model}
In this subsection, we will introduce our Traffic Accident Risk Prediction Method based on LSTM (TARPML), and Figure \ref{fig:model} illustrates the deep model of TARPML. The input layers are consisted of  two parts. The first input is the sequence of recent traffic accident frequency, and it is input to the first LSTM layer. The second input contains the longitude and latitude of the region center that we expected to predict, and it directly input into fully connected layers. The hidden layers of deep model is consisted of 4 LSTM layers and 3 fully connected layers sequentially. The last layer of model is output layer, which outputs the predicted traffic accident risk(frequency) for the given input.

To avoid overfitting, we add a dropout layer with $0.5$ dropout rate between each two fully connected layers. The activation function of fully connected layers and output layer is Rectified Linear Units (RELU), which can be denoted as $\max(0,x)$ mathematically.

\begin{figure}[!t]
\centering
\includegraphics[width=1.5 in]{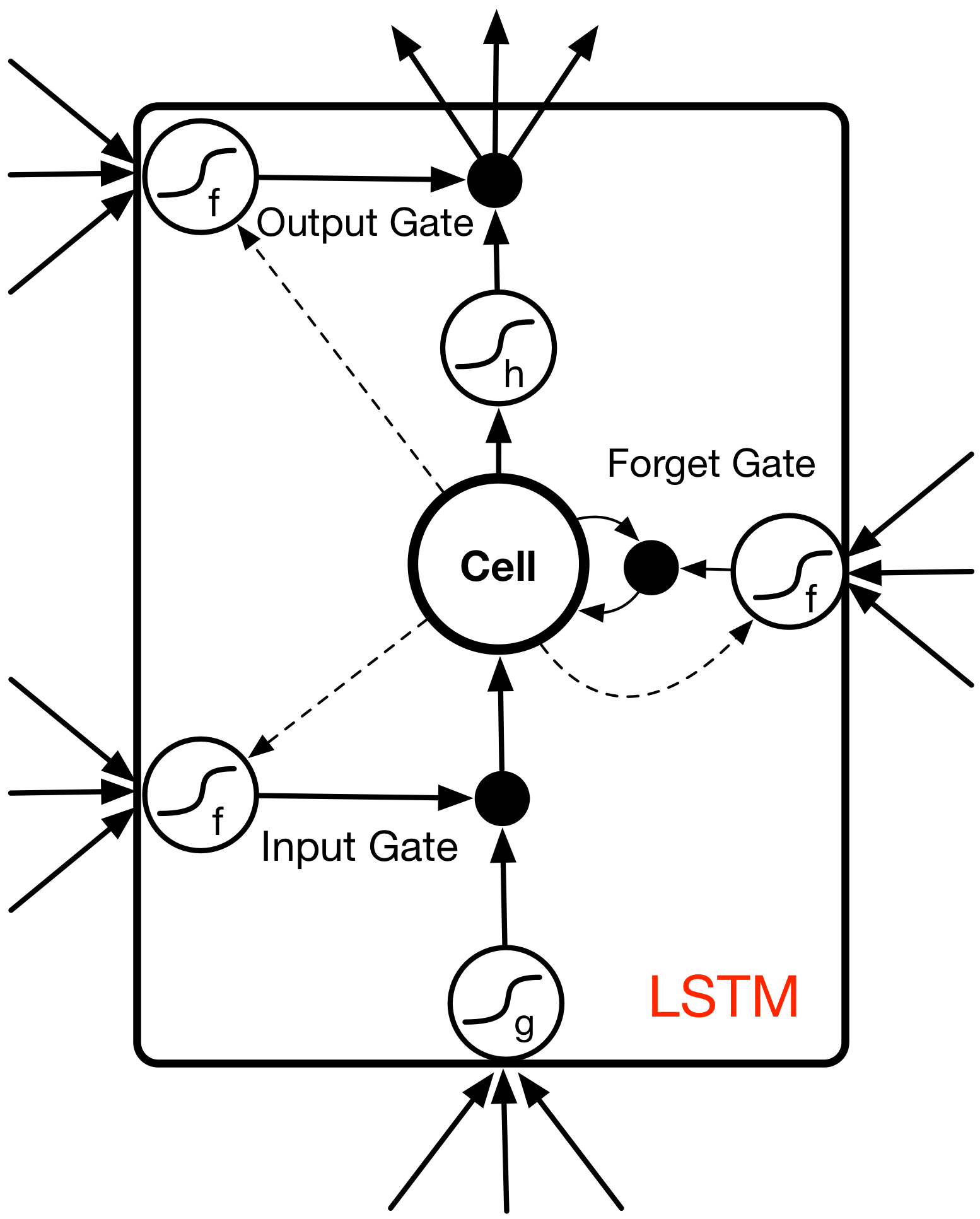}
\caption{The structure of LSTM Cell, which is consisted of an input gate, a neuron with a self-recurrent link, a forget gate and an output gate.}
\label{fig:rnn}
\end{figure}

The reason why we chose LSTM is that LSTM can capture the periodic feature of traffic accident, and traditional RNNs shows poor performance and intrinsic difficulties in training when it has long time period. These weaknesses have been proved in researches related with traffic flow prediction\cite{ma2015large}. On another hand, the explicit memory cell in LSTM can avoid the problems of gradient vanish or gradient explosion existed in traditional RNNs. The structure of LSTM is similar to traditional RNNs, and it consisted of one input layer, one or several hidden layer and one output layer. The core concept of LSTM is its memory cell in hidden layer, it contains 4 major parts: an input gate, a neuron with a self-recurrent link, a forget gate and an output gate, and its inner structure is shown in Figure \ref{fig:rnn}.

\begin{figure*}[t]
\centering
\includegraphics[width=6.5 in]{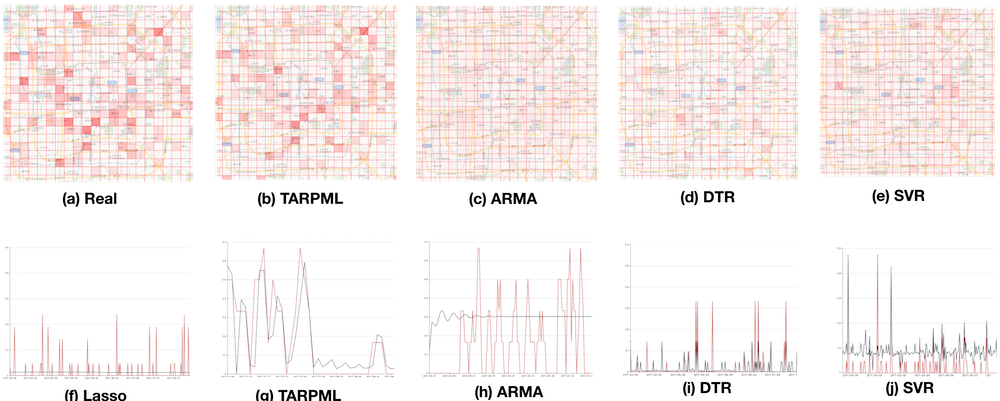}
\caption{Comparison of the real traffic accident risk map (a) and the predicted results of different models (b) - (e). (f) - (j) are the predicted risk curve from different models and its corresponding real traffic risk curve.}
\label{fig5}
\end{figure*}

\section{Experiments and Results}
In this section, we compare our TARPML method with several baseline models, including Lasso, Support Vector Regression, Random Forest Regression, etc. All of the experiments are performed by a PC (CPU: Intel Xeon(R) CPU E5-2609, 32GB memory, GPU: Tesla K20C).

\subsection{Experimental Setup}

Because our model is temporal related, we arrange the data chronologically. We chose the data from 2016-01-01 to 2017-04-01 as the training data, and the data from 2017-04-01 to 2017-08-20 is for testing. The last 20\% of training data is used as validation data. The sample size of training, validation, testing is 1590958, 397740 and 233850 respectively.

The architecture of TARPML are built upon Keras, which is a Python Deep Learning library. We chose mean squared error as objective function of optimization, and selected RMSProp as the optimizer.

By comparing the Root Mean Square Error (RMSE) of TARPML method with different input sequence length (Table \ref{rmse_of_seqlen}), we finally chose 100 as the best sequence length to input. The number of neurons of the LSTM layers are 100, 200, 200, 200, respectively, and the number of neurons of each hidden layer is 200.

\begin{table}[htbp]
\centering
\caption{Performance Comparasion of TARPML method with different input sequence lengths}
\label{rmse_of_seqlen}
\begin{tabular}{|c|c|c|c|c|}
\hline
\begin{tabular}[c]{@{}c@{}}Sequence \\ Length\end{tabular} & 10 & 20 & 50 & 100 \\ \hline
1 day & 0.119 & 0.122 & 0.115 & 0.105 \\ \hline
3 days & 0.042 & 0.041 & 0.038 & 0.034 \\ \hline
7 days & 0.022 & 0.021 & 0.018 & 0.015 \\ \hline
30 days & 0.006 & 0.005 & 0.005 & 0.004 \\ \hline
\end{tabular}
\end{table}

\subsection{Performance Evaluation}

\subsubsection{Evaluation Metrics}
To evaluate the accuracy and precision of the prediction, we selected Mean Absolute Error (MAE), Mean Squared Error  (MSE) and Root Mean Squared Error  (RMSE) as our metrics. They are defined as:

\begin{equation}
MAE = \frac{1}{n} \sum_{i=1}^n |r_i - \widehat{r_i}|
\label{eq:mae}
\end{equation}

\begin{equation}
MSE = \frac{1}{n} \sum_{i=1}^n(r_i - \widehat{r_i})^2
\label{eq:mre}
\end{equation}

\begin{equation}
RMSE = \sqrt{\frac{1}{n} \sum_{i=1}^n(r_i - \widehat{r_i})^2}
\label{eq:rmse}
\end{equation}
where $n$ is the sample size, $r_i$ and $\widehat{r_i}$ are real and predicted risk (accident frequency) respectively.

\subsubsection{Baseline Models}
We selected several traditional machine learning models as our baseline models to compare the prediction performance with our TARPML method. The baseline models we selected are Lasso, Support Vector Regression (SVR), Decision Tree Regression (DTR) and Autoregressive Moving Average Model (ARMA) . All these models were implemented by scikit-learn, a Python module that implemented lots of state-of-the-art machine learning algorithms, and the default parameters of baseline models were used. 

\subsubsection{Performance Evaluation}
We compared the predictability of our model with that of baselines, and Table \ref{performance} demonstrates their MAE, MRE and RMSE values for 3-day traffic accident frequency (Its definition can be found at Section \ref{sec:deep_model}). The table shows that our model outperforms than other models, and have less prediction errors.

\begin{table}[!t]
\setlength{\abovecaptionskip}{0pt}
\setlength{\belowcaptionskip}{5pt}
\centering
\caption{Performance for 3-day Traffic Accident Frequency Prediction}
\label{performance}
\begin{tabular}{p{2.0cm}<{\centering}p{1.2cm}<{\centering}p{1.2cm}<{\centering}p{1.2cm}<{\centering}}
\hline
Method & MAE & MSE & RMSE \\ \hline
Lasso        & 0.046   & 0.006   & 0.076    \\ \hline
SVR        & 0.066   & 0.006   & 0.075    \\ \hline
DTR       & 0.021   & 0.004   & 0.058    \\ \hline
ARMA       & 0.058   & 0.049   & 0.169    \\ \hline
\textbf{TARPML}    & 0.014   & 0.001  & 0.034   \\ \hline
\end{tabular}
\end{table}

\subsubsection{Simulation Results}
To evaluate the effectiveness of our model, here we selected 2017-07-10 (Monday) as a example for comparing the prediction results of different models. Figure \ref{fig5} (a) - (e) are the real traffic accident risk map (a) and the predicted results of different models (b) - (e), and it can be clearly seen that TARPML model is far better than other models. Figure \ref{fig5}(f) - (j) are the predicted risk curve from different models and its corresponding real traffic risk curve. It can also observed that the predicted curve from TARPML model are more accurate than others.

\section{Conclusion}

In this paper, we collected big traffic accident data, and built a deep learning model based on LSTM for predicting traffic accident risk. Based on the pattern analysis result, it can be observed that the traffic accident risk are not uniformly distributed in space and time. It shows strong periodical temporal patterns and regional spatial correlation. According to the dimensionality reduction result (t-SNE, Figure \ref{fig_tsne}), it is hard to predict traffic accident directly. Therefore, we defined the traffic accident risk by its frequency, and built a deep learning model based on LSTM to capture its spatial and temporal patterns. The performance comparison based on RMSE (Table \ref{performance}) and the predicted risk map (Figure \ref{fig5} )shows the accuracy and effectiveness of our model. This study therefore indicates that benefits gained from temporal-spatial features, big traffic accident data and deep recurrent neural network can bring accurate traffic accident risk prediction. Our method can be easily applied to the traffic accident warning system and help people avoiding traffic accident by choosing safer regions.

However, due to the complexity of traffic accident, our study has some limitations in following aspects. First, here we only utilized the traffic accident data itself for prediction. However, other related data, such as traffic flow, human mobility, road characteristic and special events, maybe significant to traffic accident risk prediction as well. Second, our prediction results are coarse-grained, and can not provide road level accident risk prediction. But it can be easily applied to the road network based prediction. Therefore, future work combined with structure of urban road network and comprehensive factors related with traffic accident will be promising to make better prediction result.

\bibliographystyle{IEEEtran}
\bibliography{ref} 
\end{document}